\documentstyle[prl,aps,12pt]{revtex}

\begin{document} \title{Fringe Patterns of Bose Condensates} \author{S.
M\'etens, D. Lima, P. Borckmans, G. Dewel\\
 Service de Chimie-Physique and \\ Center for Nonlinear Phenomena and
Complex Systems\\ C.P. 231, Universit\'e Libre de Bruxelles, \\ 1050
Bruxelles, Belgium} \date{\today} \maketitle \begin{abstract} 
We investigate within the Gross-Pitaevski (GP) theory the formation of
fringe patterns between two Bose condensates stored in a double well
potential modelling the forces applied to the experimental systems.  In
the case of repulsive interactions between the atoms, we report the
onset of interference structures similar to those observed in the
experiments, after the release of the confining potential.  Conversely,
attractive interactions lead to the collapse of the condensate when the
number of particles is larger than a critical value.  We show that a
bias field introduced in the center of the trap allows the system to
avoid the blow-up of the density and gives rise to a periodic behavior
of growth and decay of spatial modulations reminiscent of the
Fermi-Pasta-Ulam recurrence.  \end{abstract} 

\indent

The successful experimental realizations
\cite{Anderson,Bradley,Bradley2,Davis} of Bose-Einstein condensation
(BEC) in   dilute alkali atomic vapors have sparkled a renewed interest
in the study of weakly coupled Bose gases. At temperatures close to
zero, these magnetically trapped Bose condensed clouds are well
described by the condensate wavefunction $\Psi({\bf r},t)$ which obeys
the GP equation initially introduced for superfluid helium
\cite{Pitaevski1}

\begin{eqnarray}
    i\hbar \frac{\partial \Psi}{\partial t}= [
			-\frac{\hbar^2}{2m}\nabla^2 +U|\Psi|^2
			+V_{ext}({\bf r})] \Psi,          \label{nlse}
\end{eqnarray}

\noindent with the normalization condition

\begin{eqnarray}
 \int |\Psi({\bf r},t)|^2 d{ \bf r} = N,                 \label{cons}
\end{eqnarray}

\noindent where $N$ is the number of particles in the condensate.  The
interactions between atoms of mass $m$ is described by the potential
$U$. Since the thermal wavelength is larger than the characteristic
length associated with binary collisions, this potential can simply be
characterized by the $s$- wave scattering length:
\mbox{$U=4\pi\hbar^2a/m$}.  The interaction is repulsive \mbox{($a>0$)}
for \mbox{$^{87}Rb$} and \mbox{$^{23}Na$} and attractive
\mbox{($a<0$)} for \mbox{$^{7}Li$.} The confining potential is given by
\mbox{$V_{ext}({\bf r})$}.  Many works have been devoted to the
characterization of the ground state properties and collective
excitations of such condensed gases confined by a single harmonic
potential \cite{Ruprecht,Baym,Dalfovo,Stringari}.

Recently interference fringes have been observed  to form when two
condensates of repulsive atoms are allowed to expand and overlap
\cite{Andrews}. Interference between bose condensates with $a>0$ has
also been the subject of various theoretical studies
\cite{Hoston,Naraschewski,Wallis}.
 In this letter, we investigate the formation of such fringe patterns
stressing the difference between the cases of attractive and repulsive
interaction and we report
 new dynamical behaviors for bose gases with $a<0$. In this framework,
we numerically solve the GP equation to follow the time evolution of
two condensates with an equal number of atoms initially stored in a
double well potential. It consists of the superposition of a harmonic
potential describing the magnetic trap and a gaussian barrier in the
center modelling the laser sheet
 that cuts the condensate in half.  As in the experiments, the
iso-potentials of the harmonic trap are ellipsoidal with their long
axis defining vertical direction $z$.  The ratio \mbox{ ${\Omega =
\frac{ \omega_z }{ \omega_{\perp} } < 1 }$} of the corresponding
angular frequencies of the trap in the vertical and transverse
directions characterizes the anisotropy of the potential.  The height
of the barrier is chosen to make tunneling ineffective.
 This initial condition can thus be written in the form:

\begin{eqnarray}
    \Psi(z, x, y, 0) =  \phi_L(z+d, x, y, 0) + \phi_R(z-d, x, y,
0)e^{i\delta},
					\label{inicon}
\end{eqnarray}

\noindent       and  $\delta$ is the relative phase which proceeds from
the breaking of the groupe symmetry and varies between different
experimental runs.  The \mbox{$\phi$}'s are ground state wavefunctions
in each of  the two asymmetric traps created by the laser.

We first consider the case of repulsive interaction.  When $N$ is
sufficiently large (Thomas-Fermi regime), the wavefunctions are then
essentially determined by the balance between the particle interaction
energy and the external potential as the contribution of the kinetic
energy is negligible. The spatial extension \mbox{$L_0$} of $\phi$ is
much smaller than the initial separation $2d$ between the  two
condensates (Fig. 1a).  When the double trap is switched off, the
condensates expand ballistically and fringe patterns appear in the
overlap region. This expansion is essentially dominated by the
dispersive effects generated by the kinetic energy term that has been
frustrated in the ground state $\phi$ of the trap. Similar numerical
expansion dynamics are indeed obtained for $a=0$ and $a\neq 0$. The
wavelength of these interference structures increases linearly with
time $\lambda (t) =\frac{\hbar \pi t}{m d}$ and the corresponding
expansion velocity $\hbar \pi / m d $ decreases for increasing values
of the width of the gaussian barrier. The experimental fringe period
also becomes smaller for larger powers of the argon laser sheet and
thus for larger initial distances between the condensates
\cite{Andrews}. The essential dynamics can already be captured by $1D$
numerical simulations along the $z$ direction that produce the time
development illustrated in Fig. 1a which presents strong analogies with
the time-resolved diffraction patterns obtained in double-slit atomic
experiments \cite{Kurtsiefer}. These $1D$ structures are described by
the interference term $I(z,t)$ in the expression of the atomic density
$|\psi({\bf r}, t)|^2$ which takes the form:

\begin{eqnarray} I \propto \cos(\frac{ 2 \pi z}{\lambda (t)} + \phi).
\end{eqnarray}

\noindent Variation of the relative phase $\delta$ only leads to a
trivial shift of the whole structure \cite{Hoston}. Taking advantage of
the rotational symmetry about the vertical axis, two-dimensional
simulations of Eq. (1), that corresponds for instance to a section
along the $O_{xz}$ plane of the $3D$ system produce fringe patterns
(Fig. 1b) similar to those seen in the experiments on $^{23}Na$
\cite{Andrews}.  These structures unambiguously demonstrate the
coherence of those Bose condensate gas. The system chooses a phase for
the macroscopic wavefunction in a process which is analogous to the
appearance of coherent oscillations in a laser or a chemical reactor.
Indeed the GP  equation also corresponds to the strong dispersion limit
of the complex Ginzburg-Landau equation which describes these symmetry
breaking bifurcations \cite{Cross}. The atomic cloud of $5\  10^5$
atoms then behaves like a single entity to which the de Broglie
relation can be applied.

The dynamics is completely different in the case of attractive
potentials
 \mbox{$U=-|U|<0$.} This arises because
  the uniform condensate wavefunction \mbox{$\Psi_0=\sqrt{n_0} e^{i |U|
n_0 t / \hbar}$},
 an exact solution of the (GP) equation in the absence of a confining
potential,  may undergo a Benjamin-Feir instability \cite{Benjamin}
($n_0$ is the number density in the condensate). Owing to the form of
the dispersion relation

\begin{eqnarray}
 \omega_k= \frac{\hbar}{2m} \sqrt{k^4-\frac{4m n_0 |U|k^2}{\hbar^2}} ,
\end{eqnarray}

\noindent the state $\Psi_0$ is unstable to long wavelength density
perturbations with a wavenumber lying in the range

\begin{eqnarray} 0 < k < \frac{2 \sqrt{mn_0 |U|} }{ \hbar }
\end{eqnarray}

It had therefore been claimed that BEC of attractive bosons was
impossible as this condensation would have corresponded to a
mechanically unstable state for which the compressibility would become
negative \cite{Bogolubov}.
 Nevertheless convincing experimental evidence for BEC in
 a gas of \mbox{$^7Li$} atoms in a trap, consisting of permanent
magnets \cite{Tollett},  has
 been reported recently \cite{Bradley}.  In such a confined system, BEC
can be achieved when the number of particles is sufficiently small to
inhibit the destabilizing long wavelength perturbations.  More
precisely this occur when
 \mbox{$N<N_c=\alpha (l/|a|)$} where $l$ is the typical extension of
the
 ground state in the trap and $\alpha$ is a number of order unity that
 is determined by the detailed form of the confining potential
\cite{Ruprecht}.  When $N>N_c$, the attractive potential overwhelms the
zero point energy leading to the collapse of the wavefunction by which,
in 2D and 3D,  a singularity tends to form in a finite time
\cite{Kuznetsov,Pitaevski2}. A sufficient criterion for the collapse
(VPT criterion) also yields the critical number $N_c$.  In the
collapsing cloud, new effects not contained in Eq. 1 should be taken
into account such as various intrinsic inelastic processes which lead
to the heating of the sample.

In order to study the interaction between two Bose condensates of
attractive atoms, we consider the same configuration as preceedingly.
Here however we maintain some confinement at all times to preserve the
integrity of the condensed phase.  Initially two stable $(N<N_c)$
independent condensates, located near the minima at \mbox{${ z= \pm
d}$} are separated by the Gaussian barrier.  When this  barrier is
lowered a non-trivial dynamical behavior develops (Fig. 2).  During the
expansion along the z direction the unstable modes that were excluded
from the small initial condensates are awakened by the increase of the
size of the system. As a result, a structure appears that invades the
whole confining region. Contrary to the case of repulsive interactions,
the wavelength does not vary with time anymore and corresponds to that
of the fastest growing mode in the unstable range (Eq. (6)).  After
reaching a maximum value, the amplitude of the  modulated density
decreases and the system returns to a state of localized condensates
near the minima of the potential (Fig. 2).  This process repeats
periodically and its "superperiod" decreases when the intensity of the
barrier is increased.  This  behavior is reminiscent of the well
studied Fermi-Pasta-Ulam (FPU) recurrence and of other recurrences
obtained, for instance, with the nonlinear Schr\"{o}dinger equation
\cite{Yuen}, that corresponds to the $V_{ext}=0$ of Eq. ($1$).  It is
also analogous to the phenomenon of periodic alternation observed in
nonlinear optics  experiments \cite{Arecchi}.  Both phenomena consist
of a periodically ordered sequence of quasi-stationary modes.  A modal
decomposition of the wave function on the most unstable modes
reproduces the spatial periodic solutions of  the GP equation.  The
superperiod which can be calculated analytically and
 expressed in terms of elliptic functions \cite{Infeld2}  is, as the
wavelength, in good agreement with the numerical simulations.

By moving the atoms away from the center  along the z axis, the small
potential barrier acts to lower the central density hump  below the
critical value $N_c$ and so prevents the ignition of the collapse of
the global system since the cloud can be populated by more atoms before
reaching the blow-up.  The creation of a vortex in a rotating cloud can
lead to a similar effect \cite{Dalfovo}.

At larger values of $N$ (but still $N<N_c$), more unstable modes are
allowed, leading to the numerical observation of imperfect recurrences
and of
 more complex behaviors involving a restricted range of wavenumbers
\cite{Caponi}. When $N$ eventually exceeds $N_c$ each condensate
collapses separately.

The main effect of the harmonic potential is to confine the system and
consequently to reduce the number of unstable modes taking part in the
recurrence. When the anisotropy of the confining potential  is
important, $\Omega<1$, the fringes are aligned along the tranverse
direction, and the structure is periodic in the vertical direction.
When the system is confined by  a weaker potential, transverse modes
also fall into the unstable range of wavenumbers (Eq. (5)), leading to
cellular structures with spots covering the whole confinement region.
They compete with
 fringes during the recurrent motion, each one taking its turn at
dominating the solution profile.

Atomic clouds of near zero temperature, much smaller than the critical
condensation temperature, may now be obtained experimentally.  In this
regime, dissipative effects and thermal fluctuations do not affect the
macroscopic dynamics anymore. Furthermore, owing to the presence of the
trapping potential, zero temperature fluctuations may be strongly
suppressed \cite{Grossmann}.  Therefore BEC of attractive bosons could
provide an ideal candidate to  experimentally observe FPU type
recurrence phenomena.

In the presence of an initial asymmetric repartition ($n_{01}$ and
$n_{02}$)  of the atoms between the two condensates in the double trap
potential, we further report numerical observations of an oscillatory
exchange of atoms, analogous to the Josephson effect.  Numerical
simulations in both $1D$ and $2D$ exhibit the already described FPU
recurrence (Fig. 3), now accompanied by a periodic exchange of
particles locked at half the recurrence frequency.  This  locking
phenomena is observed independently of the initial density
repartition.

Contrary to the case of repulsive interactions for which the fringe
patterns can be interpreted as resulting from interferences between de
Broglie waves, condensates of attractive atoms can exhibit for
modulated initial conditions a recurrent behavior that is intimately
related to the Benjamin-Feir instability of the uniform macroscopic
wavefunction.

\acknowledgements Acknowledgements. We thank G. Nicolis for his
interest in this work and E. G. D. Cohen for fruitful discussions.  S.
M. received support from the Belgian SSTC Federal Office, (PAI
Program), D. L. from the CNPq/Brazil and P. B. and G. D. from the FNRS
(Belgium).\\

\noindent Figure 1. Interference patterns for two freely expanding Bose
condensates of \mbox{$^{23}Na$} atoms ($a= 4.9\  nm$, $N= 2\  10^5$)
initially separated by $d=40$ $\mu m$ in a double well.\\ (1a)
space-time plot of the density $|\Psi|^2$ obtained by numerically
integrating the Gross-Pitaevski equation (Eq. 1) switching off the
double-well at $t=0$.  Space is along the z direction and time runs
downward up to $t=40 ms$. The linear increase of the wavelength with
time is visible.\\ (1b) density plot in the $xz$ plane after $40$ $ms$
when the wavelength equals of $15$ $\mu m$. The gray scale interpolates
between maximum (dark) and minimum (light) densities.\\

\noindent Figure 2. Recurrent behavior of a Bose condensate of
\mbox{$^{7}Li$} atoms (\mbox{$a= -1.45\  nm$, $N=700$}) evolving in a
double well potential: sequence of density plots (plane $xz$) spanning
the recurrence " superperiod".  At time $t=0$, two identical
condensates were captured in their respective well and separated by an
adequate barrier. This barrier is lowered to produce the exhibited time
evolution.  \\

\noindent Figure 3. Josephson-type behaviour for a BE condensate of
\mbox{$^{7}Li$} atoms ($a= -1.45 nm$, $N=700$). Space-time plot of the
density along the z axis during a single exchange of particles along
the $z$ axis when the initial transients have died out.


{\small \begin{thebibliography}{Dillo83}

\bibitem{Anderson} M. H. Anderson, J. R. Ensher, M. R. Matthews, C. E.
Wieman, E. A.  Cornell, {\it Science} {\bf 269}, 198 (1995).

\bibitem{Bradley} C. C. Bradley, C. A. Sackett, R. G. Hulet, {\it Phys.
Rev. Lett.} {\bf 75}, 1687 (1995)

\bibitem{Bradley2} C. C. Bradley, C. A. Sackett, R. G. Hulet, {\it
Phys. Rev. Lett.} {\bf 78}, 985 (1997)

\bibitem{Davis} K. B. Davis, M. O. Mewes, M. R. Andrews, N. J. van
Druten, D. S. Durfee, D. M. Kurn, W. Ketterle, {\it Phys. Rev. Lett.}
{\bf 75}, 3969 (1995)


\bibitem{Pitaevski1} L. P. Pitaevski, {\it Zh. Eksp. Theor. Fiz. } {\bf
40}, 646 (1961) [{\it Sov. Phys. JETP} {\bf 13}, 451 (1961)] and E. P.
Gross, {\it Nuovo Cimento}, {\bf 20}, 454 (1961)

\bibitem{Ruprecht} P. A. Ruprecht, M. J. Holland, K. Burnett, M.
Edwards, {\it Phys. Rev. A} {\bf 51}, 4704 (1995)

\bibitem{Baym} G. Baym and C. Pethick, {\it Phys. Rev. Lett.} {\bf 76},
6 (1996)

\bibitem{Dalfovo} F. Dalfovo, L.P. Pitaevski, S. Stringari, {\it J.
Res. Natl. Inst. Stand. Technol.} {\bf 101}, 537 (1996)

\bibitem{Stringari} S. Stringari, {\it Phys. Rev. Lett.} {\bf 77}, 2360
(1996)

\bibitem{Andrews} M. R. Andrews, C. G. Towsend, H. J. Miesner, D. S.
Durfee, D. M. Kurn, W. Ketterle,   {\it Science} {\bf 275}, 637
(1997).

\bibitem{Hoston} W. Hoston, L. You, {\it Phys. Rev. A} {\bf 53}, 4254
(1996)

\bibitem{Naraschewski} M. Naraschewski et al.  {\it Phys. Rev. A} {\bf
54}, 2185 (1996)

\bibitem{Wallis} H. Wallis et al.  {\it Phys. Rev. A} {\bf 55}, 2109
(1997)

\bibitem{Kurtsiefer} C. Kurtsiefer, T. Pfau, J. Mlynek, {\it Nature}
{\bf 386}, 150 (1997)


\bibitem{Cross} M. Cross, P.C. Hohenberg, {\it Rev. Mod. Phys} {\bf
65}, 854 (1993).

\bibitem{Benjamin}T. B. Benjamin, F. E. Feir, {\it J. Fluid Mech.} {\bf
27}, 417 (1967)

\bibitem{Bogolubov} N. Bogolubov, {\it J. of Phys.} {\bf XI}, 23 (1947)

\bibitem{Tollett}J. J. Tollett, C. C. Bradley, C. A. Sackett, R. G.
Hulet, {\it Phys. Rev. A} {\bf 51}, R22 (1995)

\bibitem{Kuznetsov} E. A. Kuznetsov, J. J. Rasmussen, K. Rypdal, S. K.
Turitsyn, {\it Physica D} {\bf 87}, 273 (1995)

\bibitem{Pitaevski2} L. P. Pitaevski, {\it Phys. Lett. A} {\bf 221}, 14
(1996)

\bibitem{Yuen} H. C. Yuen, W. E. Ferguson, {\it Phys. Fluids} {\bf 21},
1275 (1978)

\bibitem{Arecchi}F. T. Arecchi, S. Boccaletti, G. B. Mindlin, C. Perez-
Garcia, {\it Phys. Rev. Lett.} {\bf 69}, 3723 (1992)

\bibitem{Infeld2} E. Infeld {\it Phys. Rev. Lett.} {\bf 47}, 717 (1981)

\bibitem{Caponi} E. A. Caponi, P. G. Saffman, H. C. Yuen, {\it Phys.
fluid} {\bf 25}, 2159 (1982)


\bibitem{Grossmann} S. Grossmann, M. Holthaus {\it Phys. Rev. Lett. }
{\bf 79}, 3557 (1957)

\end{thebibliography} }
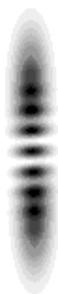
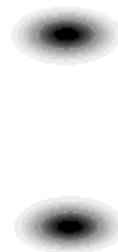
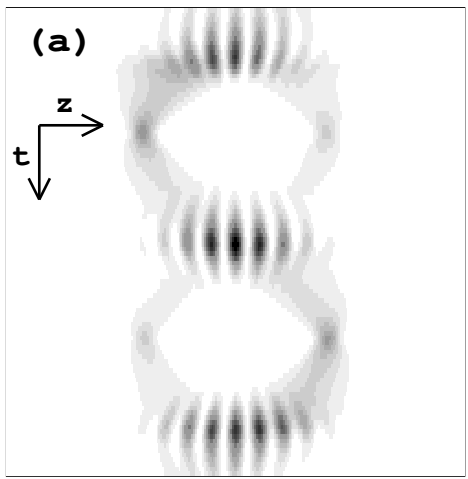
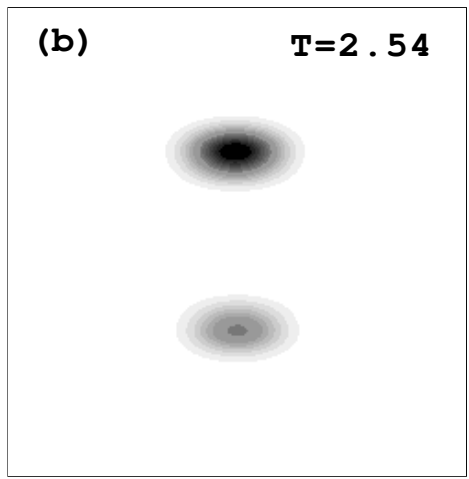
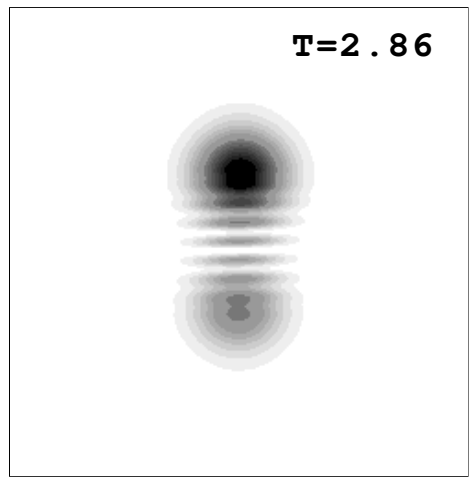
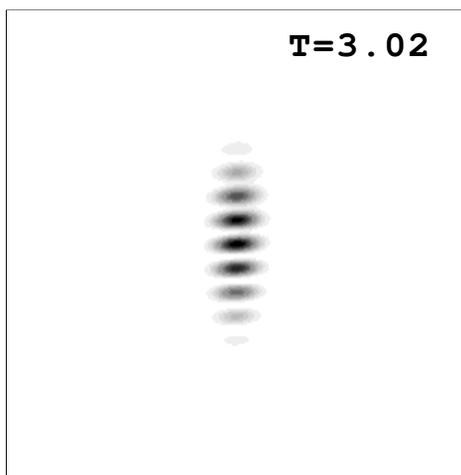
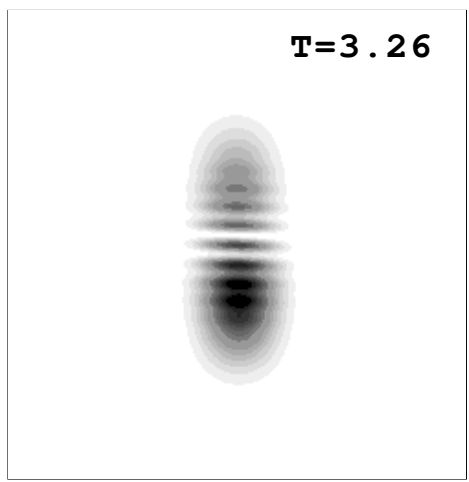
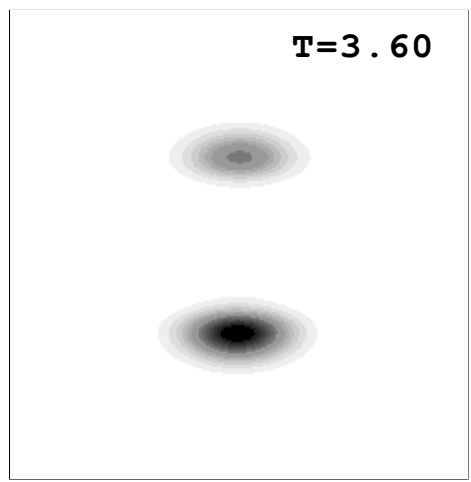
\end{document}